\documentclass[twocolumn,showpacs,preprintnumbers,amsmath,amssymb,floatfix]{revtex4}

\usepackage{graphicx}
\usepackage{dcolumn}
\usepackage{bm}
\usepackage{verbatim}
\usepackage{tabularx}

\def\be{\begin{equation}}
\def\ee{\end{equation}}
\def\bea{\begin{eqnarray}}
\def\eea{\end{eqnarray}}

\newcommand{\gsim}{ \mathop{}_{\textstyle \sim}^{\textstyle >} }
\newcommand{\lsim}{ \mathop{}_{\textstyle \sim}^{\textstyle <} }

\newcommand{\gev}{{\rm GeV}}

\newcommand{\mev}{{\rm MeV}}

\newcommand{\ccdot}{\!\cdot\!}

\begin{document}

\preprint{MCTP/08-60}

\title{Changes in Dark Matter Properties After Freeze-Out}

\author{Timothy Cohen, David E. Morrissey, and Aaron Pierce}
\affiliation{Michigan Center for Theoretical Physics,
University of Michigan, Ann Arbor MI, 48109}

\date{\today}

\begin{abstract}
The properties of the dark matter
that determine its thermal relic abundance can be very different
from the dark matter properties today.
We investigate this possibility
by coupling a dark matter sector to a scalar that undergoes
a phase transition after the dark matter freezes out.
If the value of $\Omega_\mathrm{DM}\,h^{2}$ calculated from parameters
measured at colliders and by direct and indirect detection experiments
does not match the astrophysically observed value,
a novel cosmology of this type could provide the explanation.  This mechanism also has the potential to account for the ``boost factor" required to explain the PAMELA data.
\end{abstract}

\maketitle

\section{\label{sec:Intro} Introduction}

  The amount of dark matter~(DM) in the universe has been
measured precisely by astrophysical and cosmological probes:
$(\Omega_\mathrm{DM}\,h^{2})_{astro}=0.106 \pm 0.008$ \cite{pdg}.
The leading candidate for this DM is a new weak-scale neutral particle.
If the universe follows a standard thermal history, the DM
density can be derived from measurements of the properties
of this particle.  The crucial input is the particle's thermally
averaged annihilation cross section, $\langle \sigma_{a} v \rangle$.
When the annihilation rate becomes too slow to keep pace with the
expansion of the universe the DM particle freezes out,
leaving behind a relic density of DM that merely dilutes as
the universe expands.

  With the turn-on of the Large Hadron Collider~(LHC)
and a host of direct and indirect detection experiments coming on-line,
there is hope that the nature of the DM particle will be measured
thoroughly enough that $\langle \sigma_{a} v \rangle$
can be computed. Then a prediction of the thermal relic
abundance, $(\Omega_\mathrm{DM}\,h^{2})_{particle}$,
can be made.
If $(\Omega_\mathrm{DM}\,h^{2})_{particle}
= (\Omega_{\mathrm{DM}}\,h^{2})_{astro}$,
this will be strong evidence that the universe has a standard thermal
history back to the DM freeze-out temperature, $T_{fo}$
(typically tens of GeV for weak-scale DM).  This would extend
the successful predictions of Big Bang Nucleosynthesis~(BBN),
which demonstrate a thermal history
of the universe only back to temperatures of several MeV.

  On the other hand, if the calculated relic density does not equal
the measured one, this will be evidence for physics beyond minimal thermal
DM.  If $(\Omega_\mathrm{DM}\,h^{2})_{particle}
< (\Omega_\mathrm{DM}\,h^{2})_{astro}$, it is possible that we have
not identified the dominant source
of DM or the DM was produced non-thermally as a decay product of another
particle~\cite{Moroi:1999zb}.
Conversely, if $(\Omega_{\mathrm{DM}}\,h^{2})_{particle}
> (\Omega_\mathrm{DM}\,h^{2})_{astro}$
the thermal relic abundance of the DM must have been diluted, perhaps by a
late production of entropy~\cite{Steinhardt:1983ia} or a modification
of the expansion history of the
universe~\cite{Kamionkowski:1990ni, Barrow:1982ei}.
In the present work, we explore a novel possibility that can
obtain either direction of this inequality:
a change in the properties of the DM itself
between $T_{fo}$ and the present. Time-dependent DM
has been considered in another context in attempts to
relate DM and Dark Energy~\cite{Anderson:1997un,Rosenfeld:2005pw}.

  Relevant changes in the attributes of the DM particle can occur
if there is a field whose vacuum expectation value (VEV) changes
during the crucial epoch between $T_{fo}$ and $\rm{BBN}$.
If this field influences the mass or couplings of the DM particle,
there can be a dramatic effect on the relic abundance one would calculate
based on the properties of the DM particle measured today.
Here we present a simple model that illustrates how this
mechanism could be realized.  We discuss some constraints on
scenarios of this type, and we study the phenomenology that
should accompany the late-time phase transitions typical of
this class of models.

\section{\label{sec:Ptrans} A late-time phase transition}

  To change the DM properties, we suppose
there is a phase transition~(PT) after $T_{fo}$~\cite{Frieman:1991tu}.
In the model considered here, this PT occurs in a new sector containing
a Standard Model (SM) singlet $P$.
We couple the PT sector to a model for the DM in the next section.
The PT will modify both the mass and couplings of the DM particle
in this model.

Rather than introducing a new field $P$, one might
instead try to modify the properties of the DM after freeze-out via
the electroweak PT. This does not work for electroweak-mass DM using
the minimal SM Higgs phase transition~\cite{Dimopoulos:1990ai}.
Unless the dynamics of this PT are modified (or the initial DM mass
is very large), the temperature of the PT is typically greater than
$T_{fo}$, and the DM properties would not be modified between
$T_{fo}$ and the present day. On the other hand, if the Higgs boson
sector is non-minimal, it is possible that the electroweak
transition temperature might be lowered substantially~(see \emph{e.g.}
\cite{Delaunay:2007wb}).

  The new singlet field $P$ is initially stabilized at the origin
in the early universe by a thermal mass
term~\cite{Dolan:1973qd,Weinberg:1974hy}.
As the universe cools, $P$ undergoes a PT at a temperature
$T_\mathrm{PT} < T_{fo} \approx m_{\mathrm{DM}}/20$, and develops
a non-zero VEV, $\langle P\rangle \equiv v_{P}$.
For the PT to have a significant effect on the DM properties
(perhaps by generating a large excursion in the DM mass
$\Delta_{m} \equiv \lambda_{{\mathrm{DM}}-P}\, v_{P}$)
typically requires $v_{P} \gg T_\mathrm{PT}$
\footnote{Here $\lambda_{{\mathrm{DM}}-P}$ is a dimensionless coupling between the DM and $P$.  This assumes fermionic DM.  For scalar DM the VEV-dependent
contribution must be even larger to make a significant change
in the mass, as the new contribution should be added in quadrature.}.

  We take the potential for $P$ to be
\begin{equation}
V_{P}(T=0) = -\frac{1}{2} |m_{P}|^2 P^{2} + \frac{\lambda}{4!} P^{4},
\label{eqn:ptL}
\end{equation}
which induces
\begin{equation}\label{eq:vP}
v_P(T=0) = \sqrt{6\,|m_P|^2/ {\lambda}}
\end{equation}
below $T_\mathrm{PT}$.
The $\mathbb{Z}_2$ symmetry of this potential ($P\to -P$)
means there is a danger of forming
domain walls.  We can retain the form of the potential
while avoiding domain walls by softly breaking the
$\mathbb{Z}_2$ with a very small cubic term, making
this symmetry only approximate~{\cite{Abel:1995wk}}.

  A large hierarchy between $v_{P}$ and $T_\mathrm{PT}$ in this
scenario requires that the coupling responsible for inducing a
thermal mass for $P$ be considerably larger than $\lambda$.
This can arise if $P$ couples to other states that are
approximately massless when $v_P = 0$. Such states can
emerge if $P$ is part of a larger ``hidden'' sector, perhaps coupled
to the SM only via a ``Higgs
portal''~\cite{Patt:2006fw,Schabinger:2005ei}.  For concreteness,
we consider additional fermionic fields coupling to $P$ according to
$\mathcal{L} \ni \lambda_{PQ_i} P\,\overline{Q_{i}}\,Q_{i}$. Since
the $Q$'s have no other mass terms (which would violate the
$\mathbb{Z}_2$ of $P$), these couplings contribute to the
temperature-dependent mass of the $P$ field, strongly trapping it at
the origin.  When $v_P$ shifts to its non-zero value, the $Q$'s
acquire a mass of $\lambda_{PQ_i}\,v_P$, typically of order a few hundred GeV.

  At high temperatures and near the origin of $P$, the potential
is approximately~\cite{Dolan:1973qd, Weinberg:1974hy}
\begin{equation}
V_{P}(T) = -\frac{1}{2}(|m_P|^2-\frac{N_Q}{6}\lambda_{PQ}^2\,T^2)P^2
+\frac{1}{4!}\lambda\,P^4,
\end{equation}
where $\lambda_{PQ}$ is the (universal) coupling between $P$ and the $Q$'s
and $N_Q$ is the number of Dirac $Q$ fields.
This potential gives a PT temperature of
\begin{equation}\label{eq:TPT}
T_\mathrm{PT} = \sqrt{\frac{{6}|m_P|^2}{{N_Q}\,\lambda_{PQ}^2}}.
\end{equation}

  Strong trapping of the $P$ field at the origin typically leads
to a brief period of thermal inflation~(TI)~{\cite{Lyth:1995ka,
Yamamoto:1985rd, Lazarides:1985ja}}.
The vacuum energy density during TI is $\rho_{vac} =|m_P|^2 v_P^2/4$.
If TI ends at $T_\mathrm{PT}$ by the instantaneous decay of
the $P$ field to radiation, we can estimate the reheating
temperature $T_\mathrm{RH}$ via conservation of energy:
\begin{eqnarray}
T_\mathrm{RH}^4 &=& \frac{45\,|m_P|^4}{g_*^\mathrm{RH}\,\pi^2\,\lambda}
+ \frac{{36}\,g_*^\mathrm{PT}\,|m_P|^4}{g_*^\mathrm{RH}\,
{N_Q}^2\,\lambda_{PQ}^4},
\label{eq:T_RH}
\end{eqnarray}
where $g_*$ is the effective number of relativistic degrees of freedom.
Reheating can dilute the DM abundance.  Although this is not the dominant
effect that we wish to explore, it can be of quantitative importance.
This is also the reason why we rely on thermal corrections,
rather than an additional cubic term in the tree-level potential,
to trap $P$ at the origin.  With a cubic term, the trapping need not
turn off as the universe supercools and could lead to a severe
dilution of the DM abundance.

  To estimate this dilution, we first assume there are no new sources
of entropy during TI.  This fixes
${n_{fo}}/{s_{fo}}= {n_\mathrm{PT}}/{s_\mathrm{PT}}$,
where $n$ and $s$ are the number density of the DM and entropy density
of the universe respectively.
No DM is produced in the reheating process,
implying $n_\mathrm{PT} = n_\mathrm{RH}$.
Once TI ends and reheating completes, the new conserved quantity is
$n_\mathrm{RH}/s_\mathrm{RH} = n_\mathrm{PT}/s_\mathrm{RH}$.
The dilution factor, $D$, is
\begin{equation}\label{eq:dilution}
\frac{n_\mathrm{PT}}{s_\mathrm{RH}} =
\frac{s_\mathrm{PT}}{s_\mathrm{RH}}\frac{n_{fo}}{s_{fo}}
= \left(\frac{g_*^\mathrm{PT}}{g_*^\mathrm{RH}}\frac{T_\mathrm{PT}^3}{T_\mathrm{RH}^3}\right)
\frac{n_{fo}}{s_{fo}} \equiv  D\times\frac{n_{fo}}{s_{fo}}.
\end{equation}
Taking into account the change in the mass of the particle,
the present abundance is given by
\begin{equation}\label{eq:relicDensity}
(\Omega_\mathrm{DM}\,h^2)_{astro}  =
D\times \left(\frac{m_\mathrm{DM}^{v_P \ne 0}}{m_\mathrm{DM}^{v_P = 0}}\right)
\times \Omega^{v_P = 0}_\mathrm{DM}h^2.
\end{equation}
This can differ dramatically from $(\Omega_\mathrm{DM}\,h^{2})_{particle}$,
as we will see in the next section.

  In Table~\ref{tab:paramsForPT} we exhibit a benchmark point
that gives a first-order PT with a transition temperature
$T_\mathrm{PT}\ll v_P$.
To obtain this feature, the value of $\lambda$ is small.
This interaction obtains additive corrections of the form
$\Delta\lambda = \sum(c_{b} \, \lambda_{b}^2- c_{f} \,
\lambda_{f}^4)/(16\,\pi^2)$,
where the sum runs over bosons and fermions that couple to $P$,
and the $c_{i}$ are ${\mathcal O}(1)$ coefficients.
For the benchmark couplings, the small value of $\lambda$
is technically natural.

  The value of $T_\mathrm{PT}$ for the benchmark point is also
large and could exceed a typical value of $T_{fo}$
unless the mass of the DM particle is many hundreds of $\mbox{GeV}$.
Smaller values of $T_\mathrm{PT}$ can be achieved by reducing the value of
$|m_P|^2$.  This leads to light excitations of $P$ that can be
phenomenologically problematic -- it is difficult to make
them decay quickly enough to avoid BBN constraints while not disturbing
the evolution of supernovae.

\begin{table}[ttt]
\begin{center}
\begin{tabular}{||>{\centering\arraybackslash} m{.134\columnwidth}
                 |>{\centering\arraybackslash} m{.16\columnwidth}
                 |>{\centering\arraybackslash} m{.134\columnwidth}
                 |>{\centering\arraybackslash} m{.065\columnwidth}
                 |>{\centering\arraybackslash} m{.055\columnwidth}
                 |>{\centering\arraybackslash} m{.125\columnwidth}
                 |>{\centering\arraybackslash} m{.125\columnwidth}
                 |>{\centering\arraybackslash} m{.06\columnwidth}||}
\hline
\hline
$|m_P|$ & $\lambda$  & $v_{P}$ & $\lambda_{PQ}$  & $N_Q$ & $T_{\mathrm{PT}}$ & $T_{\mathrm{RH}}$ & $D$ \\
\hline
4.0 GeV & $1.5\times 10^{-5}$ & 2.5 TeV & 0.10 & 9 & 33 GeV & 40 GeV & 0.77 \\
\hline
\hline
\end{tabular}
\caption{Benchmark phase transition parameters.}
\label{tab:paramsForPT}
\end{center}
\vspace{-0.1cm}
\end{table}

\section{\label{sec:Dmtoy} A Dark Matter sector}

  There are many possibilities for the DM sector,
all of which could work with the generic phase transition module
we presented in the previous section.  The particular DM sector
we consider is a ``level-changing'' model, consisting of three fermions with
the same quantum numbers as the Higgsinos and Bino
of the minimal supersymmetric SM:
a vector-like pair of $SU(2)_L$ doublets $\psi_L$ and $\psi_{\bar{L}}$
with the appropriate hypercharges, and a gauge singlet $\psi_{S}$.
All fields in this DM sector are charged
under an exact $X\to -X$ symmetry (independent of the approximate
$\mathbb{Z}_2$ of $P$), implying that the lightest of these
particles is absolutely stable.  The DM sector Lagrangian is
\begin{eqnarray}
\mathcal{L} &\ni& \mu\,\psi_L\ccdot\psi_{\bar{L}}
+\lambda_1\, H\ccdot\psi_L\,\psi_s
+\lambda_2\,H^{*}\ccdot\psi_{\bar{L}} \,\psi_s\\
&&
+\,(\mu_s+\lambda_s\, P) \, \psi_s\,\psi_s + \mathrm{h.c.}, \nonumber
\end{eqnarray}
where $H = (G^+,\frac{1}{\sqrt{2}}(H^0+ i \,G^0))^T$ is the SM Higgs boson.
The resulting ``neutralino'' mass matrix is
\begin{eqnarray}\label{eq:DMMassMatrix}
\mathcal{M}^0 = \left( \begin{array}{ccc}
             0 & \mu & -\lambda_1\,\frac{v_H}{\sqrt{2}}\\
             \mu & 0  & \lambda_2\,\frac{v_H}{\sqrt{2}} \\
             -\lambda_1\,\frac{v_H}{\sqrt{2}} & \lambda_2\,\frac{v_H}{\sqrt{2}} & 2\,(\mu_s+\lambda_s\,v_P)
             \end{array} \right),
\end{eqnarray}
with electroweak VEV $\langle H^0 \rangle \equiv v_H = 246\,\gev$ \footnote{
The $\psi_s\psi_s$
coupling breaks the approximate $\mathbb{Z}_2$
symmetry of $P$, and thus quantum corrections from loops of the $\psi_s$
field would modify the $P$ potential in Eq.\,\eqref{eqn:ptL}.
This can easily be avoided by adding a second singlet
(or another pair of doublets)
without significantly altering the DM story we present here.
To avoid complication, we will consider only one singlet.
These symmetries could also forbid a $v_P$ dependent ``Higgsino"
mass which we also ignore for simplicity.}.

\newcolumntype{S}{>{\centering\arraybackslash} m{.1822\columnwidth} }
\newcolumntype{T}{>{\centering\arraybackslash} m{.2323\columnwidth} }
\begin{table}[ttt]
\begin{center}
\begin{tabular}{||S|S|S|S|S||}
\hline
\hline
$\mu$ & $\mu_s$ & $\lambda_{s}$ & $\lambda_1$ & $\lambda_2$\\
\hline
1.3 TeV & 0.68 TeV & -0.070 & 0.020 & 0.010 \\
\hline
\end{tabular}
\begin{tabular}{||T|T|T|T||}
\hline
$m_\mathrm{DM}(v_P = 0)$ & $m_\mathrm{DM}(v_P \ne 0)$ & $T_{fo}(v_P = 0)$ & $T_{fo}(v_P \ne 0)$  \\
\hline
1.3 TeV & 1.0 TeV & 65 GeV & 52 GeV\\
\hline
\hline
\end{tabular}
\caption{Benchmark parameters realizing $(\Omega_\mathrm{DM}\,h^2)_{particle}>(\Omega_\mathrm{DM}\,h^2)_{astro}$.}
\label{tab:paramsForDM}
\end{center}
\vspace{-0.1cm}
\end{table}

  Within this model, it is not difficult to obtain
$(\Omega_\mathrm{DM}\,h^{2})_{particle}
\gg (\Omega_\mathrm{DM}\,h^{2})_{astro}$.
As an example, we consider the benchmark parameter point
given in Tables~\ref{tab:paramsForPT} and \ref{tab:paramsForDM}.
At high temperatures $v_{P}=0$. There the DM is a nearly pure
combination of the doublets $\psi_{L}$ and $\psi_{\bar{L}}$:
$X^{0}   \approx  1/\sqrt{2} \, \psi_{L}
+ 1/\sqrt{2} \, \psi_{\bar{L}} + \epsilon \, \psi_{s},$
with $\epsilon \approx (\lambda_1-\lambda_2)v_H/(4\,\mu_s-2\,\mu)$.
The thermal relic abundance of
this state is nearly identical to that of a pure Higgsino.
This is set by its annihilation to pairs of $W$ bosons,
and is given by~\cite{Mahbubani:2005pt}
\begin{equation}\label{eq:omegaHiggsino}
\Omega_\mathrm{DM}^{v_P=0}\,h^2 =
0.1 \left(\frac{m_\mathrm{DM}}{1\,\mathrm{TeV}}\right)^2,
\end{equation}
including coannihilation with the heavier ``charginos''.

The mass and composition of the DM change after the PT.
For the parameters in the Tables, the lightest of the DM-sector particles
is nearly pure singlet post-PT.  Using Eq.\,(\ref{eq:relicDensity}), its relic
density is $(\Omega_\mathrm{DM}\,h^2)_{astro} = 0.1$.  This is the value
measured by astrophysical probes.  However, it is considerably
different from the value one would reconstruct from measurements
of the DM particle Lagrangian today, assuming one measured
the relevant couplings but did not take into account the non-canonical
cosmological effect described here~\footnote{
In practice, it is difficult to measure
the couplings of this particular DM candidate.
However, we see no fundamental impediment to building models with
DM candidates amenable to experimental study.}.

  The dominant contribution to the \emph{apparent} particle
annihilation cross section, assuming the relevant particles and
their couplings can be measured, is the $s$-channel exchange of a
$P$ going into $Q\,\overline{Q}$. Assuming a standard thermal history,
the predicted relic density is approximately given by
\begin{equation}
(\Omega_\mathrm{DM}\,h^2)_{particle}
= \frac{0.02}
{N_{Q} \, (\lambda_{PQ} \, \lambda_{s})^{2}}
\left(\frac{m_\mathrm{DM}}{1\,\mathrm{TeV}}\right)^2,
\end{equation}
yielding $(\Omega_\mathrm{DM}\,h^2)_{particle}  = 45$ for the
benchmark, more then two orders of magnitude larger than
$(\Omega_\mathrm{DM}\,h^2)_{astro}$. Even if the
PT-sector particles are not discovered at colliders, the properties
of the DM today will differ from those at freeze-out.  These
properties can potentially still be deduced by direct and indirect
detection searches for DM.

  We obtained $(\Omega_\mathrm{DM}\,h^{2})_{particle}
\gg (\Omega_\mathrm{DM}\,h^{2})_{astro}$ in this example.
A different choice of mass matrix (Eq. (\ref{eq:DMMassMatrix})) can lead to the opposite relationship.
When this is the case, the value of
$\langle \sigma_a v \rangle$ should increase after the PT,
and the DM can potentially recouple after
thermal inflation.  Demanding that the DM stay frozen out
gives a bound on the allowed change in the relic density.
Non-recoupling of the DM after reheating requires
\begin{eqnarray}\label{eq:recouplingAtRH}
n_\mathrm{PT} \, \langle
\sigma_a\, v \rangle^{v_P\neq0}
&\leq& 1.66 (g_*^\mathrm{RH})^{1/2} \frac{T_\mathrm{RH}^2}{M_\mathrm{Pl}},
\end{eqnarray}
where $M_\mathrm{Pl}$ is the Planck mass.  A similar condition holds
for the initial ($v_P=0$) freeze-out cross section and temperature.
Combining these expressions and accounting for
redshift from freeze-out to the PT gives
\begin{equation}\label{eq:conditionOnXSection}
\frac{ {\langle \sigma_av \rangle}^{v_P\neq 0} }
{ {\langle \sigma_a\, v \rangle}^{v_P=0} } \leq
\frac{ \sqrt{g_*^\mathrm{RH} g_*^{v_P=0}} } {g_*^\mathrm{PT}}
\left( \frac{T_\mathrm{RH}^{2} \, T_{fo}^{v_P=0}}
{T_{\mathrm{PT}}^{3}} \right).
\end{equation}
Here $g_{*}^{v_{P}=0}$  is the effective number of relativistic
degrees of freedom calculated at $T_{fo}^{v_{P} =0}$.
Using the standard approximate solution to the
Boltzmann equation \cite{Kolb:1990vq} to relate
$\langle \sigma_av \rangle$ to
$\Omega_\mathrm{DM}\,h^2$, along with Eq.~(\ref{eq:relicDensity}),
leads to the constraint
\begin{equation}\label{eq:conditionOnOmega}
\frac{(\Omega_\mathrm{DM}\,h^2)_{particle}}{(\Omega_\mathrm{DM}\,h^2)_{astro}}
\gsim
\sqrt{\frac{g_*^\mathrm{RH}}{g_*^{v_P\neq 0}}}
\frac{T_\mathrm{RH}}{T_{fo}^{v_P\neq 0}}.
\end{equation}

A large change in the apparent relic density
without recoupling requires a hierarchy
between $T_\mathrm{RH}$ and $T_{fo}^{v_P\neq 0}$.  To avoid disturbing
BBN, $T_\mathrm{RH}$ must be larger than about $10\,\mev$.
Taking a typical $T_{fo}$ of tens of GeV, the apparent
relic density can be reduced by a factor of a thousand.
In practice we find it difficult to obtain such low reheating temperatures
simultaneous with the large $v_P$ needed to make a significant shift
in the DM properties.

\section{\label{sec:Pheno} Phenomenology of a Late Phase Transition}

  For the PT to happen after DM freeze-out, the mass of
the physical $P$ excitation $\sim |m_P|$ should be light.
The existence of a light $P$ is the most
generic feature of the mechanism presented here, and so it is worth
considering its phenomenology in some detail.
The symmetries of the model allow the Lagrangian term
$\mathcal{L} \ni (\lambda_{PH}/2)P^{2}\, |H|^{2}$,
coupling $P$ with the SM Higgs boson.  The resultant mixing with
the Higgs boson gives two mass eigenstates, $p^0$ and $h^0$.
The mixing angle is given by
\begin{equation}
\tan2\theta=\frac{6\,\lambda_{PH}\,v_P\,v_H}
{\lambda_Hv_H^2-\lambda\,v_P^2},
\end{equation}
where ${\mathcal L} \ni \lambda_{H} (H^{0})^{4}/4!$ \footnote{With a non-zero $\lambda_{PH}$ term, the VEV of $P$ modifies the potential for $H^{0}$ and vice versa. Therefore, the Higgs VEV can be different when $v_{P} = 0$ (changing the mass of the $W$ boson).  It is also important to check that this cross coupling still allows a well separated $T_{\mathrm{EW}} > T_{\mathrm{PT}}$.}.

  The relevant phenomenological constraints and signals depend
on the precise mass of the $p^0$, which in principle could range
from tens of GeV all the way down to a fraction of an MeV.  Mixing allows the
$p^0$ to be produced in association with a $Z^{0}$,  or  to appear
in meson decays.  For $m_{p^0} \lsim$ 100 MeV, astrophysical constraints
similar to those for axions \cite{Raffelt} become important.

  For DM masses near the weak scale, the natural value of the $p^0$
mass is on the order of a few GeV.  For the parameters
in Table~\ref{tab:paramsForPT} and a moderate mixing angle,
$m_{p^0} \approx 6 $ GeV.  In this mass range, the $p^0$ could
be produced in Upsilon ($\Upsilon$) decays.
To lowest order \cite{Wilczek:1977zn},
\begin{equation}
\frac{\Gamma(\Upsilon\rightarrow p^0\,\gamma)}
{\Gamma(\Upsilon\rightarrow \mu^+\,\mu^-)}
=
\frac{\sin^2\theta\,m_b^2}{2\,\pi\,v_H^2\,\alpha}
\left(1-\frac{m_{p^0}^2}{m_{\Upsilon}^2}\right).
\end{equation}
Requiring $\mathrm{BR}(\Upsilon\rightarrow p^0\,\gamma)\times
\mathrm{BR}(p^0\rightarrow \tau^+\,\tau^-)\lesssim 10^{-5}$ \cite{Love:2008hs}
gives a modest bound on the mixing angle of $\theta\lesssim 0.3$
for BR$(p^0\rightarrow \tau^+\,\tau^-)=1$.
For a 6 GeV $p^0$, decays to charm quarks actually exceed those to
$\tau$'s by a factor of 2 (unless a more complicated Higgs sector
allows for a $\tan \beta$ enhanced $p^0$ couplings to down-type
fermions). In this mass range, a comparable bound exists from
non-observation of $Z^0\,p^0$, which would have been seen in
$Z^0\,h^0$ searches at LEP~\cite{Barate:2003sz}.  For higher masses,
$m_{P} > 10$ GeV, the bound on the mixing angle strengthens due to
the LEP constraint: $\theta < 0.14$. The $p^0$ decays are very
prompt; the lifetime of $p^0$ is $2 \times 10^{-19}\,\mathrm{sec}$
for $\theta = 0.14$.  The $p^0$ branching ratios are
identical to a SM Higgs boson of the same mass.  If the $p^0$ mass
falls below the $B$-meson mass, bounds on the mixing angle from $b
\rightarrow s\, P$ processes \cite{pdg} are strong: $\theta \lsim
10^{-4}$.

  These considerations also provide a way to observe
the $p^0$ state at colliders.
For lighter masses ($m_{p^0} < 8\,\gev$) and large mixing,
searches for the rare decay
$\Upsilon\rightarrow \gamma\,p^0\rightarrow\gamma\,\tau^+\tau^-$ may be useful.
If  the $h^0$ is not so heavy that it decays to $W$ bosons, then
$h^{0} \rightarrow p^{0}\, p^{0}$ need only compete with
$h^{0} \rightarrow b\, \bar{b}$~\cite{Cheung:2007sva}.  The ratio of the widths is given by
\begin{equation}
\frac{\Gamma(h^0\rightarrow p^0\,p^0)}{\Gamma(h^0\rightarrow b\,\overline{b})}
= \frac{3 \, (\xi \lambda c_{\theta}^2  v_P  v_{H})^{2}}{m_b^2}
\frac{(m_{h^0}^2-4\,m_{p^0}^2)^{1/2}}{(m_{h^0}^2-4\, m_b^2)^{3/2}}
\end{equation}
with the effective coupling
\begin{eqnarray}
\xi &=& t_{\theta}+\frac{\lambda_H\,v_H}{\lambda\,v_P}\,t_{\theta}^2 \\
&-& \frac{t_{2\,\theta}}{18} \left(1-\frac{\lambda_H\,v_H^2}{\lambda\,v_P^2}\right)\left[1-2\,t_{\theta}^2-\frac{v_P}{v_H}(2\,t_{\theta}-t_{\theta}^3)\right], \nonumber
\end{eqnarray}
where $t_{\theta} \equiv \tan \theta$ and $c_{\theta} \equiv \cos \theta$.
For $\theta$ saturating
the LEP bound, BR($h^{0} \rightarrow p^{0} p^{0}$) can
approach 40\% for $v_{P} \gsim 750$ GeV.
Then $h^0\,Z^0\rightarrow p^0\,p^0\,Z^0 \rightarrow 4\,b\, Z^{0}$
might be observable at the LHC if $b$-tagging efficiencies
are sufficiently high~\cite{Carena:2007jk}, though it will be challenging.

  Thus far we have not mentioned the decay of the $Q$'s.
This can proceed via higher-dimension operators. Alternately, the
$Q$'s can decay to quarks through renormalizable operators if they are
$SU(3)_c$ triplets and are allowed a very small mixing with the
quarks of the SM.  For the benchmark parameters in
Table~\ref{tab:paramsForPT}, $m_{Q} = 250$ GeV.  Then given the
latter scenario there is the possibility of producing the $Q$'s
directly at the LHC.

\section{\label{sec:Discussion} Discussion}

  The DM model presented here represents an existence proof
of a general mechanism: DM properties can change after freeze-out.
We have focused our attention on situations where a shifting VEV
causes a change in the DM mass and composition.
A similar effect could occur if the
coupling that sets the relic abundance of the DM is a
function of a light modulus.

  In another example, the DM mass might shift so that
$2\, m_\mathrm{DM}$ is approximately resonant with some
other state in the theory, such as a Higgs boson.
With the DM now sitting on resonance, one would calculate
a tiny thermal relic abundance.  To implement
this scenario using  a SM Higgs boson is difficult.
For a ``natural" PT, $m_\mathrm{DM}\sim \mathrm{TeV}$.
To access the Higgs resonance,
$m_{h^0}\sim2\,m_\mathrm{DM}\sim\mathcal{O}(\mathrm{TeV})$.
This implies $\Gamma_{h^0}$ will be too large to generate
a strong resonant enhancement of $\langle \sigma_a\,v\rangle$.
In the presence of heavy but
narrow resonances, this is a viable mechanism.

  Alternately, the DM itself could remain unchanged, but the properties
of particles crucial for setting the thermal relic abundance are modified
by the cosmology.
Consider a coannihilating particle, $C$, nearly degenerate with the DM.
If the mass of $C$ shifts between
$T_{fo}$ and now, the importance of coannihilation would not be evident
from low-temperature measurements, and the calculated
$(\Omega_\mathrm{DM}\,h^{2})_{particle}$ would differ from the true value.

 If $(\Omega_\mathrm{DM}\,h^{2})_{particle}
\neq (\Omega_\mathrm{DM}\,h^{2})_{astro}$,
it is possible that the relic abundance of
the DM is actually thermal, but an alternate cosmology has altered
the DM properties since freeze-out.  These scenarios are naturally
realized if there is a light modulus that undergoes a late PT.
The field responsible for the late time transition, $P$,
could show up in future experiments.  One possibility is via Higgs
boson decays: $h^0 \rightarrow p^0\,p^0$.  If $p^0$ is light enough,
it could also be produced in rare meson decays.  Embedding a model of
this type in an extension
of the minimal supersymmetric SM is a direction for future investigation.

Recent preliminary data from the PAMELA~\cite{Boezio:2008},
ATIC~\cite{Chang:2005}, and PPB-BETS~\cite{Torii:2008xu} experiments report significant excesses of cosmic
ray positrons and electrons above the expected astrophysical background.
This excess could be the result of dark matter annihilation in our galaxy.
However, such a dark matter interpretation of these results requires
a DM annihilation cross-section well above the value that would generate
the observed dark matter relic density~\cite{Grajek:2008jb}.  
The cosmology discussed here offers 
the possibility of explaining these indirect DM signals while maintaining 
a fairly standard thermal freeze-out picture.  What is needed is a 
DM annihilation cross-section that increases significantly between 
freeze-out and today.

{\it Acknowledgments:} AP acknowledges the hospitality
of the Kavli Institute of Theoretical Physics and the
Aspen Center for Physics during the completion
of this work.  We also thank Dan Chung, Paolo Gondolo, Scott Watson and Neal Weiner for discussions.
AP and TC are supported by NSF CAREER Grant NSF-PHY-0743315.
DM is supported by DOE Grant DE-FG02-95ER40899.

\bibliography{CDMF}

\end{document}